\documentclass[aps,prb,preprint,groupedaddress]{revtex4}
\usepackage{amsmath}
\usepackage{amssymb}
\usepackage{epsfig}
\usepackage{graphicx,amsmath}
\usepackage{calc}
\usepackage{bm}
\begin{document}

\title{Metal-insulator transition in a strongly-correlated two-dimensional electron system}

\author{A.~A. Shashkin$^a$ and S.~V. Kravchenko$^b$}

\affiliation{$^a$Institute of Solid State Physics, Chernogolovka, Moscow District 142432, Russia\\
$^b$Physics Department, Northeastern University, Boston, Massachusetts 02115, USA}

\begin{abstract}
Experimental results on the metal-insulator transition and related phenomena in strongly interacting two-dimensional electron systems are discussed. Special attention is given to recent results for the strongly enhanced spin susceptibility, effective mass, and thermopower in low-disordered silicon MOSFETs.
\end{abstract}
\maketitle




\section{Strongly and weakly interacting 2D electron systems}

In two-dimensional (2D) electron systems, the electrons are confined in a one-dimensional potential well and are free to move in a plane. The strongly-interacting limit is reached in such systems at low electron densities where the kinetic energy is overwhelmed by the energy of the electron-electron interactions. The interaction strength is characterized by the ratio between the Coulomb energy and the Fermi energy, $r_s^*=E_{ee}/E_F$. If we assume that the effective electron mass is equal to the band mass, the interaction parameter $r_s^*$ in the single-valley case reduces to the Wigner-Seitz radius,
\begin{equation}
r_s=1/(\pi n_s)^{1/2}a_B,
\end{equation}
and, therefore, increases as the electron density, $n_s$, decreases (here $a_B$ is the Bohr radius in semiconductor). Candidates for the ground state of the system include Wigner crystal characterized by spatial and spin ordering \cite{wigner34}, a ferromagnetic Fermi liquid with spontaneous spin ordering \cite{stoner46}, paramagnetic Fermi liquid \cite{landau57}, etc. In the strongly-interacting limit ($r_s\gg1$), no analytical theory has been developed so far. Numeric simulations \cite{tanatar89} show that the Wigner crystallization is expected in a very dilute regime, when $r_s$ reaches approximately 35. More recent refined numeric simulations \cite{attaccalite02} have predicted that prior to the Wigner crystallization, in the range of the interaction parameter $25\leq r_s\leq35$, the ground state of the system is a strongly correlated ferromagnetic Fermi liquid. At higher electron densities, $r_s\sim1$ (weakly-interacting regime), the electron liquid is expected to be paramagnetic, with the effective mass, $m$, and Land\'e $g$ factor renormalized by interactions. In addition to the ferromagnetic Fermi liquid, other intermediate phases between the Wigner crystal and the paramagnetic Fermi liquid may also exist.

In real 2D electron systems, the always-present disorder leads to a drastic change of the above simplified picture, which dramatically complicates the problem. According to the scaling theory of localization \cite{abrahams79}, in an infinite disordered noninteracting 2D system, all electrons become localized at zero temperature and zero magnetic field. At finite temperatures, regimes of strong and weak localizations can be distinguished:\\
(i) if the conductivity of the 2D electron layer has an activated character, the resistivity diverges exponentially as $T\rightarrow0$; and\\
(ii) in the opposite limit (the so-called weak localization), the resistivity increases logarithmically with decreasing temperature --- an effect originating from the increased probability of electron backscattering from impurities to the starting point. The incorporation of weak interactions ($r_s<1$) between the electrons adds to the localization \cite{altshuler80}. However, for weak disorder and $r_s\geq1$, a possible metallic ground state was predicted \cite{finkelstein83,finkelstein84,castellani84}.

In view of the competition between the interactions and disorder, one can consider high- and low-disorder limits. In highly-disordered electron systems, the range of low densities is not accessible as the strong (Anderson) localization sets in. This corresponds to the weakly-interacting limit in which an insulating ground state is expected. Much more interesting is the case of low-disordered electron systems because low electron densities corresponding to the strongly-interacting limit become accessible. According to the renormalization group analysis for multi-valley 2D systems \cite{punnoose05}, the strong electron-electron interactions can stabilize the metallic ground state, leading to the existence of a metal-insulator transition in zero magnetic field.

In this chapter, we will focus on experimental results obtained in low-disordered strongly interacting 2D electron systems; in particular, in (100)-silicon metal-oxide-semiconductor field-effect transistors (MOSFETs). Due to the relatively large effective mass, relatively small dielectric constant, and the presence of two valleys in the spectrum, the interaction parameter in silicon MOSFETs is an order of magnitude larger at the same electron densities compared to that in the 2D electron system in n-GaAs/AlGaAs heterostructures: except at extremely low electron densities, the latter electron system can be considered weakly interacting. Indeed, the observed effects of strong electron-electron interactions are more pronounced in silicon MOSFETs compared to n-GaAs/AlGaAs heterostructures, although the fractional quantum Hall effect, attributed to electron-electron interactions, has not been reliably established in silicon MOSFETs.

\section{Zero-field metal-insulator transition}
\label{zero}

As we have already mentioned, no conducting state was expected in zero magnetic field, at least in weakly-interacting 2D electron systems. Therefore, it came as a surprise when the behavior consistent with the existence of a metallic state and the metal-insulator transition was found in strongly-interacting 2D electron systems in silicon \cite{zavaritskaya87,kravchenko94a,kravchenko95b}.

\begin{figure}
\centerline{\psfig{file=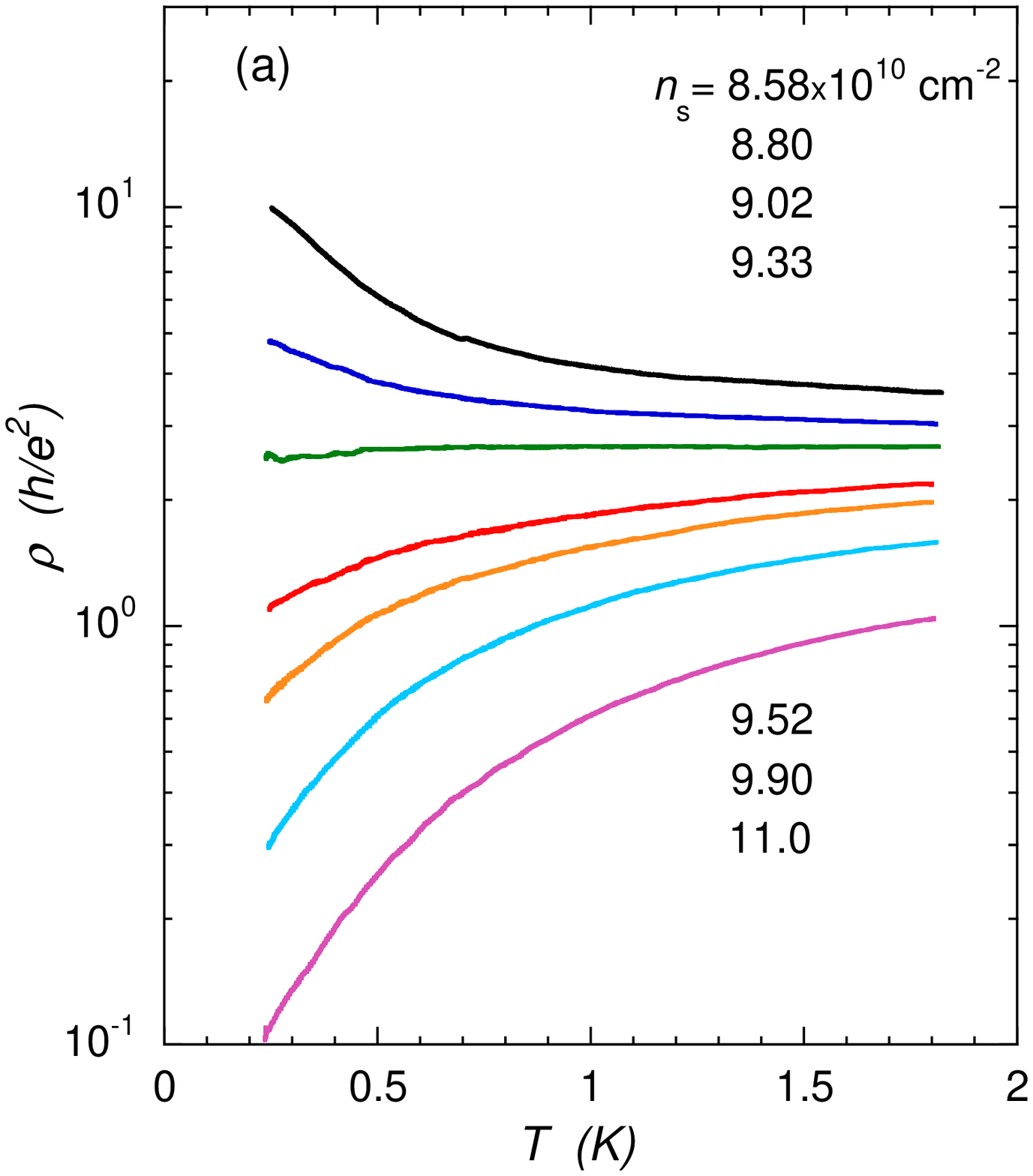,width=3.02in}\psfig{file=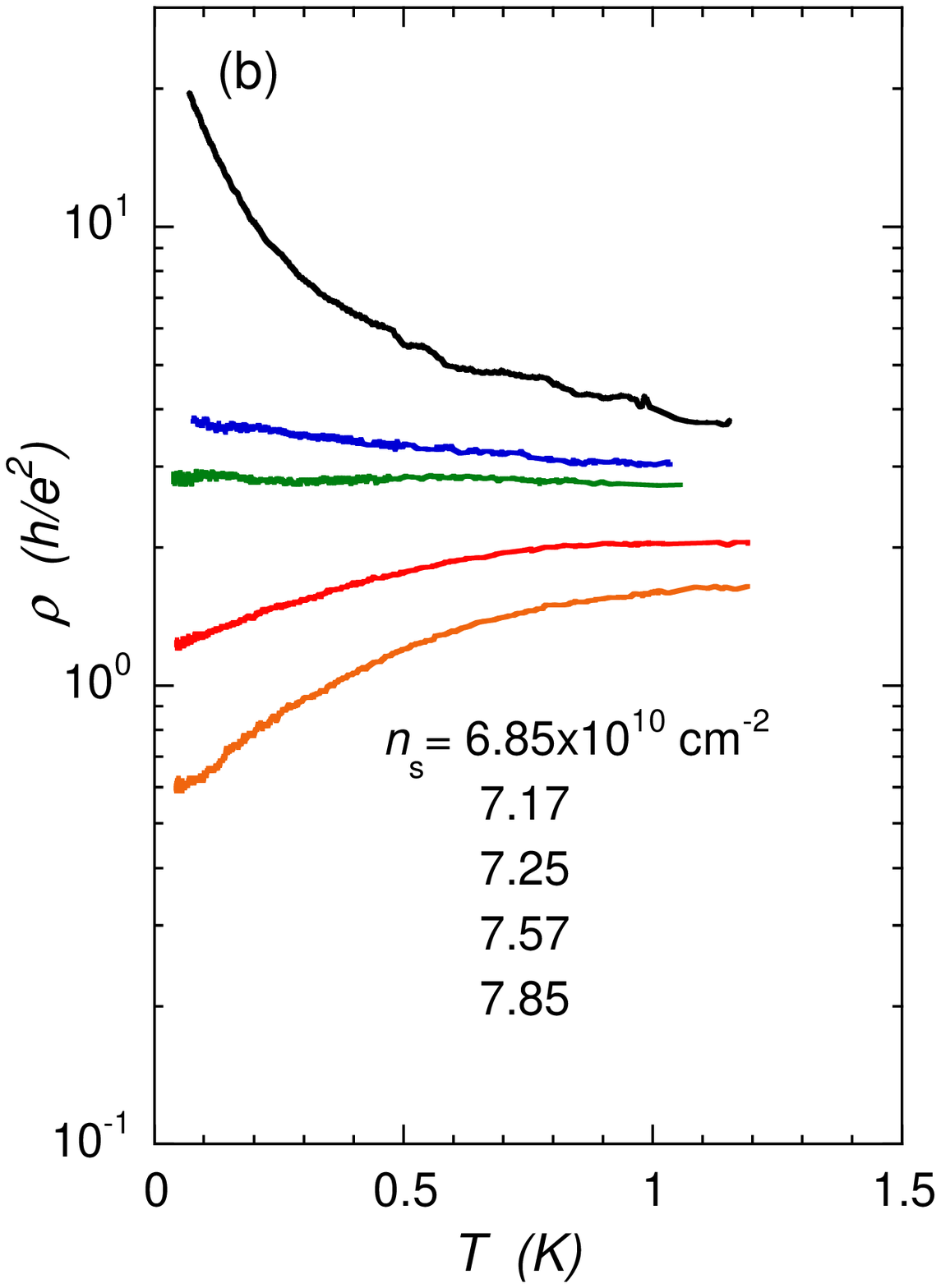,width=2.5in}}
\caption{\label{flat} The resistivity versus temperature in two Si MOSFET samples from different sources. (a)~high-mobility sample provided by V.~M. Pudalov (from Ref.~\cite{sarachik99}), and (b)~sample fabricated by R. Heemskerk and T.~M. Klapwijk (from Ref.~\cite{kravchenko00b}).}
\end{figure}

Provided that the temperature dependences of the resistance are strong, the curves with positive (negative) derivative $d\rho/dT$ are indicative of a metal (insulator) states \cite{sarachik99,abrahams01,kravchenko04}. If extrapolation of $\rho(T)$ to $T=0$ is valid, the critical point for the metal-insulator transition is given by $d\rho/dT=0$. In a low-disordered 2D electron system in silicon MOSFETs, the resistivity at a certain ``critical'' electron density shows virtually no temperature dependence over a wide range of temperatures \cite{sarachik99,kravchenko00b} (Fig.~\ref{flat}(a,b)). This curve separates those with positive and negative $d\rho/dT$ nearly symmetrically at temperatures above 0.2~K \cite{abrahams01,kravchenko04}. Assuming that it remains flat down to $T=0$, one obtains the critical point which corresponds to a resistivity $\rho\approx 3h/e^2$.

To verify whether or not the separatrix corresponds to the critical density, an independent determination of the critical point is necessary: comparison of values obtained using different criteria provides an experimental test of whether or not a true MIT exists at $B=0$. One criterion (the ``derivative criterion'') was described above; its weakness is that it requires extrapolation to zero temperature.  A second criterion can be applied based on an analysis of a temperature-independent characteristic, namely, the localization length $L$ extrapolated from the insulating phase.  These two methods have been applied to low-disordered silicon MOSFETs by Shashkin~{\it et al.} \cite{shashkin01a} and Jaroszynski~{\it et al.} \cite{jaroszynski02}.

\begin{figure}
\centerline{\psfig{file=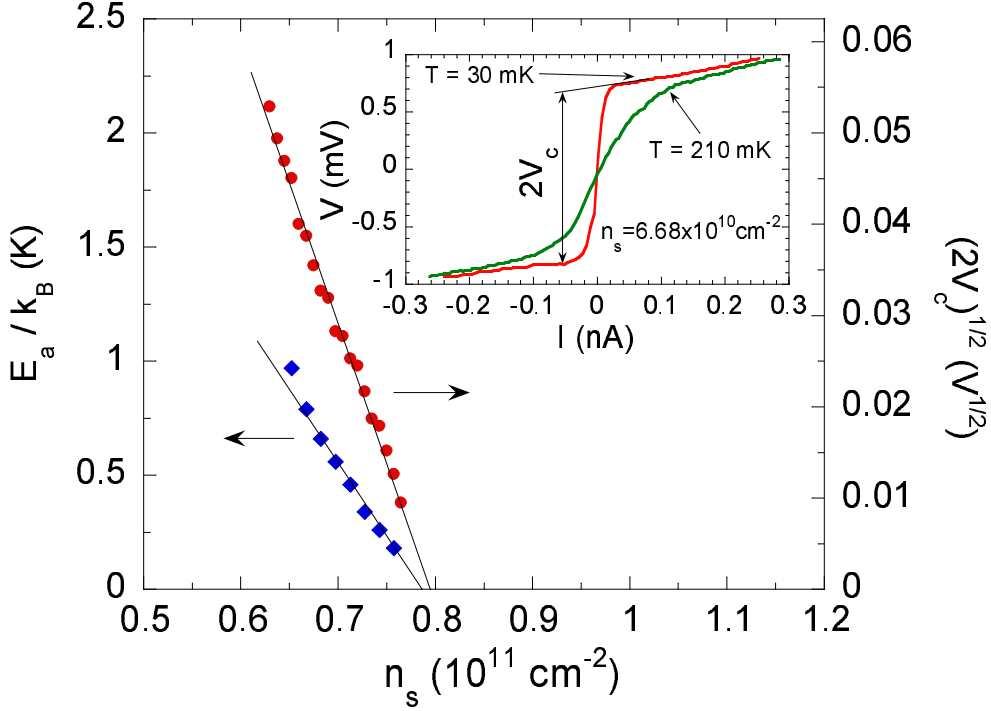,width=5in}}
\caption{\label{nc2} Activation energy (diamonds) and square root of the threshold voltage (circles) versus electron density in zero magnetic field in a low-disordered silicon MOSFET. The inset shows current-voltage characteristics recorded at $\approx30$ and $210$~mK, as labeled; note that the threshold voltage is essentially independent of temperature. From Ref.~\cite{shashkin01a}.}
\end{figure}

Deep in the insulating phase, the temperature dependence of the resistance obeys the Efros-Shklovskii variable-range hopping form \cite{mason95}; however, closer to the critical electron density and at temperatures that are not too low, the resistance has an activated form $\rho \propto e^{E_a/k_BT}$ \cite{adkins76,dolgopolov92,pudalov93,shashkin94a,shashkin94b} due to thermal activation to the mobility edge.  Fig.~\ref{nc2} shows the activation energy $E_a$ as a function of the electron density (diamonds); the data can be approximated by a linear function which yields, within the experimental uncertainty, the same critical electron density as the ``derivative criterion'' ($n_c\approx0.795\times10^{11}$~cm$^{-2}$ for the sample of Fig.~\ref{nc2}).

The critical density can also be determined by studying the nonlinear current-voltage $I-V$ characteristics on the insulating side of the transition.  A typical low-temperature $I-V$ curve is close to a step-like function: the voltage rises abruptly at low current and then saturates, as shown in the inset to Fig.~\ref{nc2}; the magnitude of the step is $2\, V_c$.  The curve becomes less sharp at higher temperatures, yet the threshold voltage, $V_c$, remains essentially unchanged.  Closer to the MIT, the threshold voltage decreases, and at $n_s=n_c\approx0.795\times10^{11}$~cm$^{-2}$, the $I-V$ curve is strictly linear \cite{shashkin01a}.  According to Refs.~\cite{polyakov93,shashkin94a,shashkin94b}, the breakdown of the localized phase occurs when the localized electrons at the Fermi level gain enough energy to reach the mobility edge in an electric field, $V_c/d$, over a distance given by the localization length, $L$, which is temperature-independent:
$eV_c(n_s)\; L(n_s)/d=E_a(n_s)$
(here $d$ is the distance between the potential probes). The dependence of $V_c^{1/2}(n_s)$ near the MIT is linear, as shown in Fig.~\ref{nc2} by closed circles, and its extrapolation to zero threshold value again yields approximately the same critical electron density as the previous criteria. The linear dependence $V_c^{1/2}(n_s)$, accompanied by linear $E_a(n_s)$, signals the localization length diverging near the critical density: $L(n_s)\propto1/(n_c-n_s)$.

These experiments indicate that in low-disordered samples, the two methods --- one based on extrapolation of $\rho(T)$ to zero temperature and a second based on the behavior of the temperature-independent localization length --- yield the same critical electron density $n_c$. This implies that the separatrix remains ``flat'' (or extrapolates to a finite resistivity) at zero temperature. Since one of the methods is independent of temperature, this equivalence supports the existence of a true $T=0$ MIT in low-disordered samples in zero magnetic field.

Additional confirmation in favor of zero-temperature zero-field metal-insulator transition is provided by magnetic \cite{shashkin05} and thermopower \cite{mokashi12} measurements, as described in next sections. We will argue that the metal-insulator transition in silicon samples with very low level of disorder is driven by interactions.

\section{Possible ferromagnetic transition}

In 2000, it was experimentally found that the ratio between the spin and the cyclotron splittings in silicon MOSFETs strongly increases at low electron densities \cite{kravchenko00a}. The spin splitting is proportional to the $g$-factor while the cyclotron splitting is inversely proportional to the effective mass; therefore, their ratio is determined by the product $g^*m^*$ which is in turn proportional to the spin susceptibility (here $m^*$ is the renormalized effective mass and $g^*$ is the renormalized g-factor). The strong enhancement of the spin susceptibility has indicated that at low electron densities, the system behaves well beyond the weakly interacting Fermi liquid.

\begin{figure}
\centerline{\psfig{file=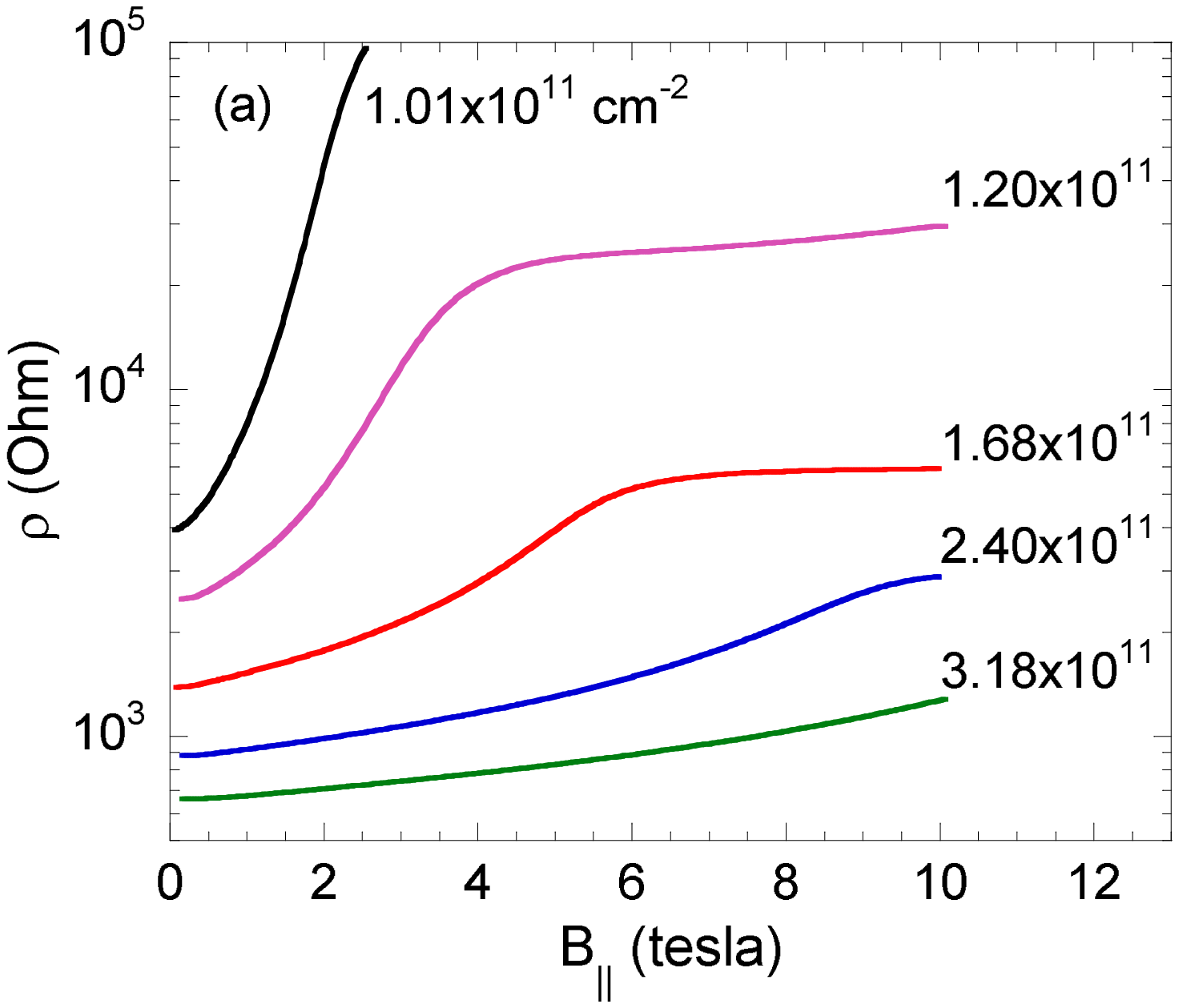,width=2.85in}\psfig{file=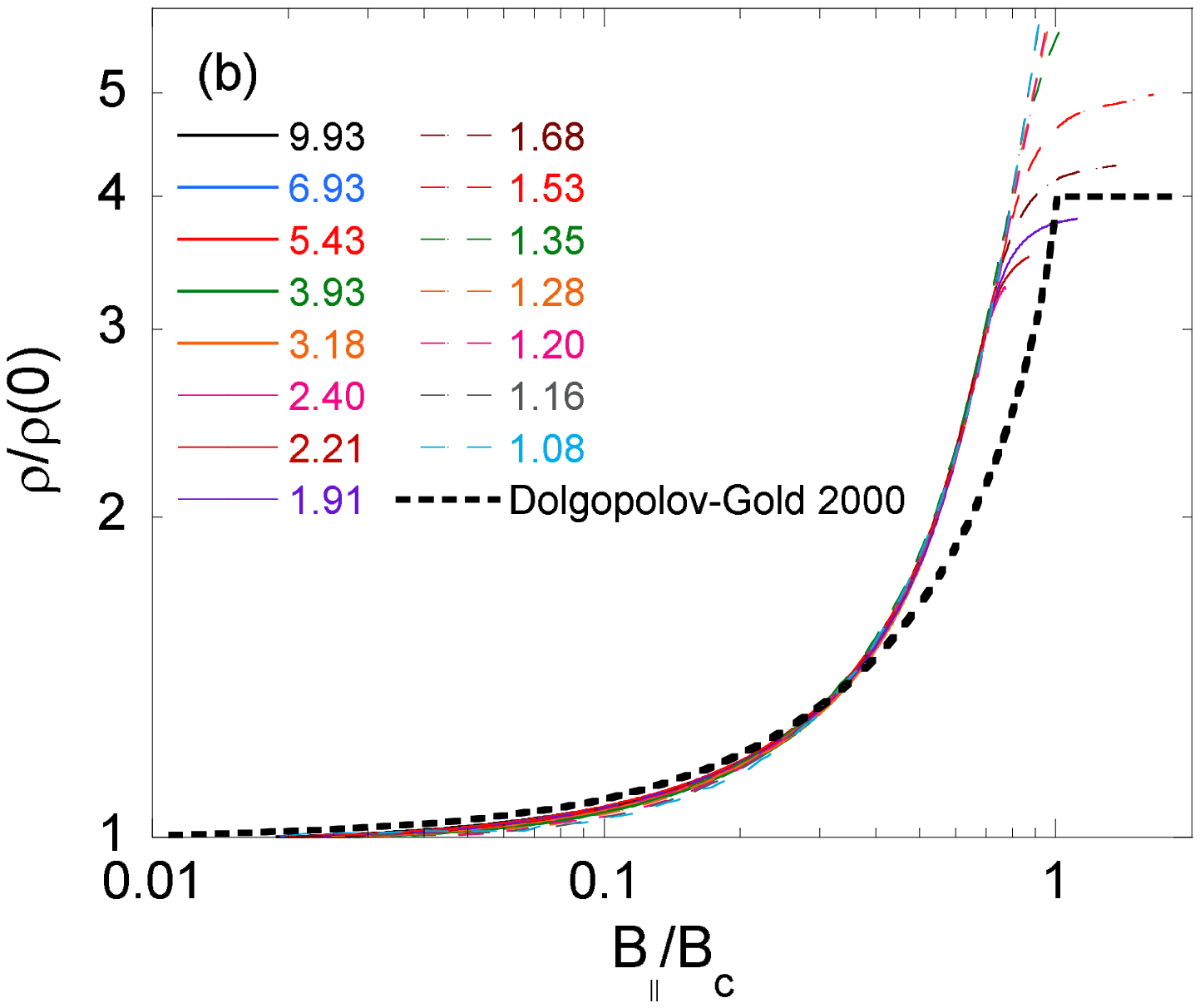,width=2.7in}}
\caption{\label{ferro} (a)~Low-temperature magnetoresistance of a clean silicon MOSFET in parallel magnetic field at different electron densities above $n_c$.  (b)~Scaled curves of the normalized magnetoresistance versus $B_\parallel/B_c$.  The electron densities are indicated in units of $10^{11}$~cm$^{-2}$.  Also shown by a thick dashed line is the normalized magnetoresistance calculated by \cite{dolgopolov00}. Adapted from Ref.~\cite{shashkin01b}.}
\end{figure}

\begin{figure}
\centerline{\psfig{file=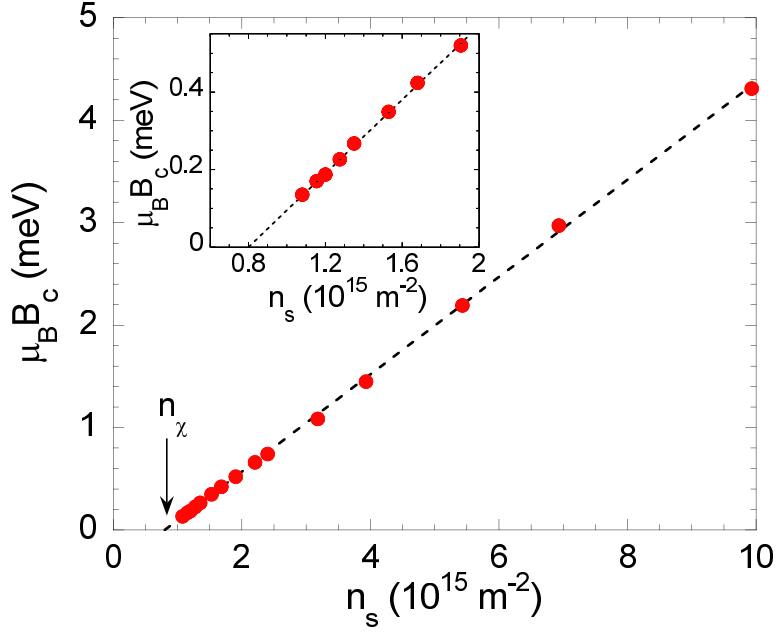,width=4in}}
\caption{\label{polarization} Scaling parameter $B_c$ (corresponding to the field required for full spin polarization) as a function of the electron density.  An expanded view of the region near $n_\chi$ is displayed in the inset.  Adapted from Ref.~\cite{shashkin01b}.}
\end{figure}

Application of a magnetic field parallel to the 2D plane promotes a strong positive magnetoresistance which, however, saturates at a certain density-dependent value of the field, $B_c$.  This saturation has been shown to correspond to the onset of the full spin polarization \cite{okamoto99,vitkalov00}; therefore, the product $g^*m^*$ can be recalculated from the parallel-field magnetoresistance data.  Shashkin {\it et al.} \cite{shashkin01b} scaled the magnetoresistivity in the spirit of the theory \cite{dolgopolov00} which predicted that at $T=0$, the normalized magnetoresistance is a universal function of the degree of spin polarization,
\begin{equation}
P\equiv g^*\mu_BB_\parallel/2E_F=g^*m^*\mu_BB_\parallel/\pi\hbar^2n_s
\end{equation}
(here, the two-fold valley degeneracy in silicon has been taken into account). Shashkin {\it et al.} \cite{shashkin01b} scaled the data obtained in the limit of very low temperatures where the magnetoresistance becomes temperature-independent and, therefore, can be considered to be at its $T=0$ value.  In this regime, the normalized magnetoresistance, $\rho(B_\parallel)/\rho(0)$, measured at different electron densities, collapses onto a single curve when plotted as a function of $B_\parallel/B_c$ (here $B_c$ is the scaling parameter, normalized to correspond to the magnetic field $B_{\rm sat}$ at which the magnetoresistance saturates).  An example of how $\rho(B_\parallel)$, plotted in Fig.~\ref{ferro}~(a), can be scaled onto a universal curve is shown in Fig.~\ref{ferro}~(b).  The resulting function is described reasonably well by the theoretical dependence predicted by Dolgopolov and Gold.  The quality of the scaling is remarkably good for $B_\parallel/B_c\le 0.7$ in the electron density range $1.08\times10^{11}$ to $10\times10^{11}$~cm$^{-2}$.  As shown in Fig.~\ref{polarization}, the scaling parameter is proportional over a wide range of electron densities to the deviation of the electron density from its critical value: $B_c\propto(n_s-n_\chi)$.

The fact that the parallel magnetic field required to produce complete spin polarization, $B_c\propto n_s/g^*m^*$, tends to vanish at a finite electron density $n_\chi\approx 8\times 10^{10}$~cm$^{-2}$ (which is close to the critical density $n_c$ for the metal-insulator transition in this electron system) points to a sharp increase of the spin susceptibility, $\chi\propto g^*m^*$, and possible ferromagnetic instability in dilute silicon MOSFETs.

\begin{figure}
\centerline{\psfig{file=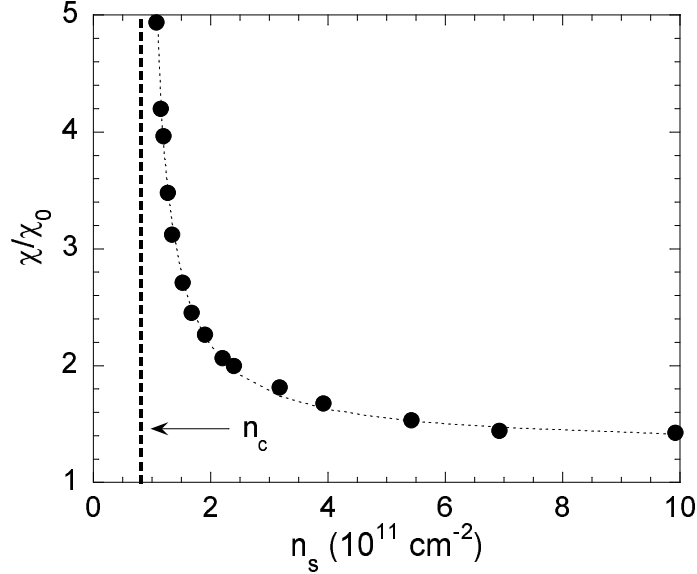,width=4in}}
\caption{\label{gm} The product $g^*m^*$ versus electron density obtained from the data for $B_c$. Critical electron density obtained from transport measurements is indicated by the dashed line. From Ref.~\cite{shashkin01b}.}
\end{figure}

The spin susceptibility, $\chi$, can be calculated using the above data. In the clean limit, the magnetic field required to fully polarize the electron spins is related to the $g$-factor and the effective mass by the equation
\begin{equation}
g^*\mu_BB_c=2E_F=\pi\hbar^2n_s/m^*.
\end{equation}
Therefore, the spin susceptibility, normalized by its ``non-interacting'' value, can be calculated as
\begin{equation}
\frac{\chi}{\chi_0}=\frac{g^*m^*}{g_0m_b}= \frac{\pi\hbar^2n_s}{2\mu_BB_cm_b}.
\end{equation}
The results of this recalculation are shown in Fig.~\ref{gm}.  One can see that the spin susceptibility increases by a factor of five compared to its non-renormalized value.

\section{Effective mass or $g$-factor?}

\begin{figure}
\centerline{\psfig{file=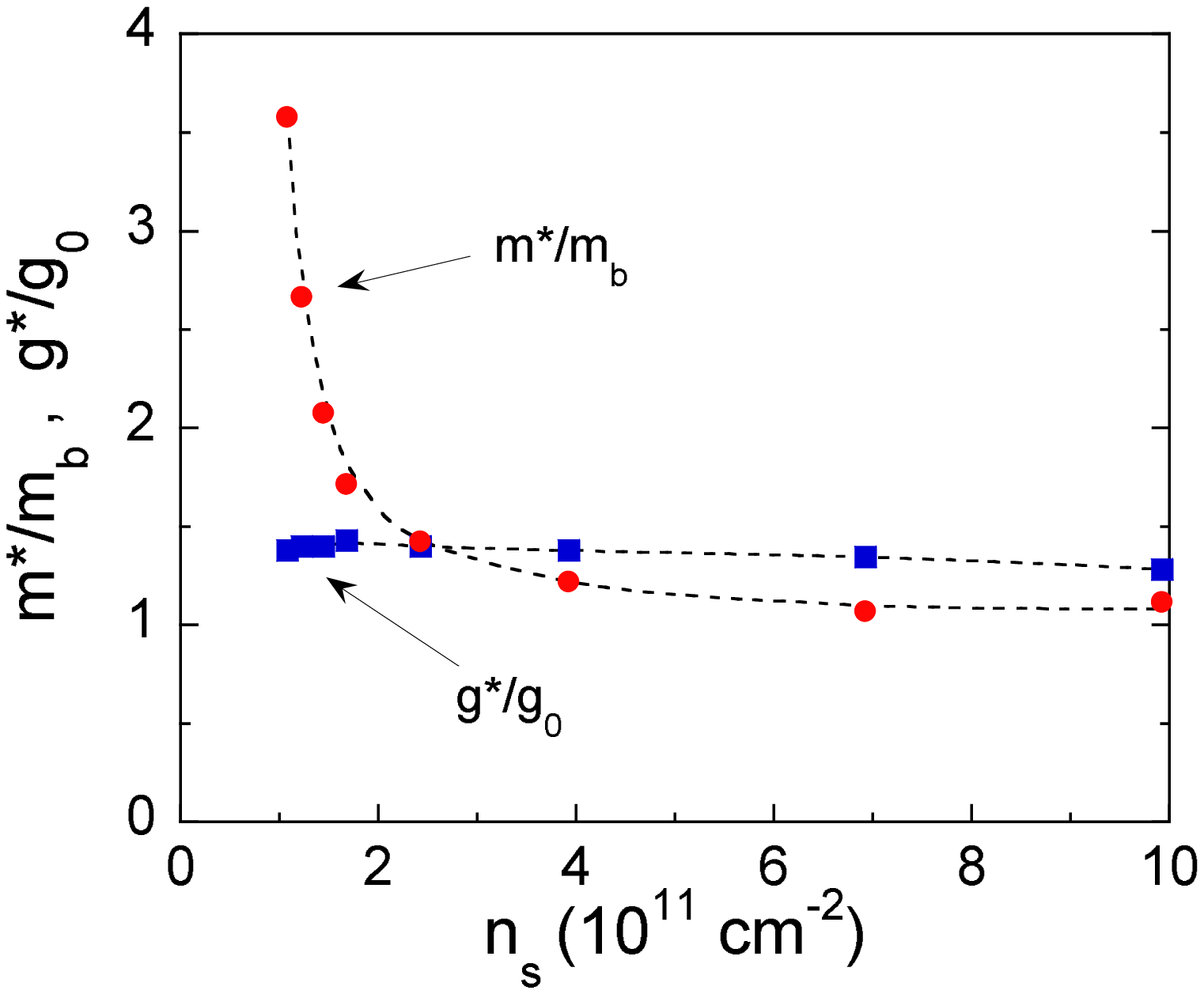,width=3in}\psfig{file=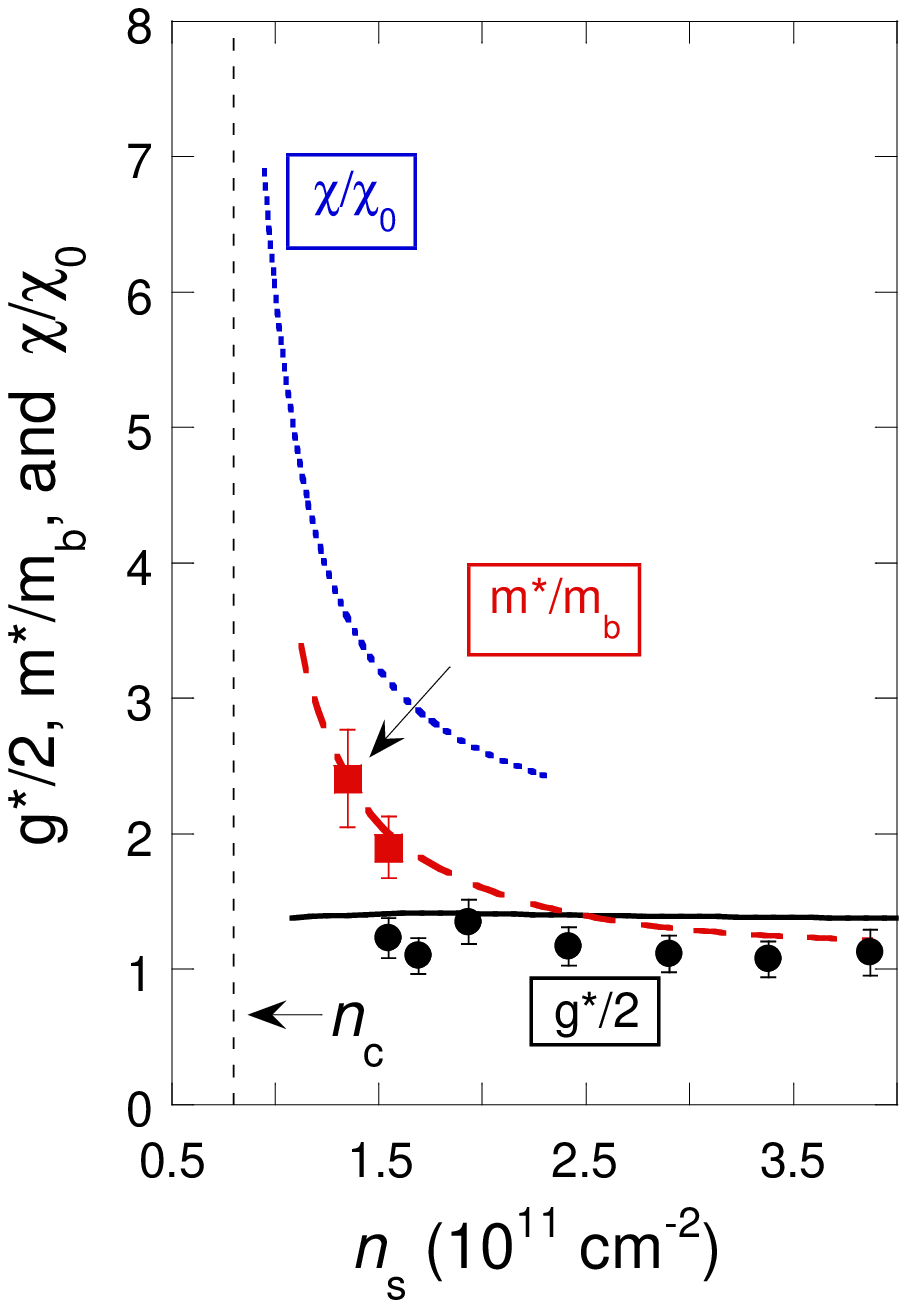,width=2.4in}}
\caption{\label{sigma} Left-hand panel: The effective mass and $g$ factor versus electron density determined from an analysis of the temperature-dependent conductivity and parallel-field magnetoresistance. The dashed lines are guides to the eye. From Ref.~\cite{shashkin02}. Right-hand panel: The effective mass (squares) and $g$ factor (circles) as a function of the electron density obtained by magnetization measurements in perpendicular and tilted magnetic fields. The solid and long-dashed lines represent, respectively, the $g$ factor and effective mass, previously obtained from transport measurements \cite{shashkin02}, and the dotted line is the Pauli spin susceptibility obtained by magnetization measurements in parallel magnetic fields \cite{shashkin06b}. The critical density $n_c$ for the metal-insulator transition is indicated. From Ref.~\cite{anissimova06}.}
\end{figure}

In principle, the increase of the spin susceptibility could be due to an enhancement of either $g^*$ or $m^*$ (or both).  Shashkin~{\it et al.} \cite{shashkin02} analyzed the data for the temperature dependence of the conductivity in zero magnetic field in spirit of the theory \cite{zala01}.  It turned out that it is the effective mass, rather than the $g$ factor, that sharply increases at low electron densities \cite{shashkin02} (Fig.~\ref{sigma}(left-hand panel)). It was found that the magnitude of the mass does not depend on the degree of spin polarization, indicating a spin-independent origin of the effective mass enhancement \cite{shashkin03a}. It was also found that the relative mass enhancement is system- and disorder-independent and is determined by electron-electron interactions only \cite{shashkin07}.

In addition to transport measurements, thermodynamic measurements of the magnetocapacitance and magnetization of a 2D electron system in low-disordered silicon MOSFETs were performed, and very similar results for the spin susceptibility, effective mass, and $g$ factor were obtained \cite{khrapai03a,shashkin06b,anissimova06} (Fig.~\ref{sigma}(right-hand panel)). The Pauli spin susceptibility behaves critically close to the critical density $n_c$ for the $B=0$ metal-insulator transition: $\chi\propto n_s/(n_s-n_\chi)$. This is in favor of the occurrence of a spontaneous spin polarization (either Wigner crystal or ferromagnetic liquid) at low $n_s$, although in currently available samples, the residual disorder conceals the origin of the low-density phase. The effective mass increases sharply with decreasing density while the enhancement of the $g$ factor is weak and practically independent of $n_s$. Unlike in the Stoner scenario, it is the effective mass that is responsible for the dramatically enhanced spin susceptibility at low electron densities.

Yet another way to estimate the behavior of the effective mass is to measure the thermoelectric power. The thermopower is defined as the ratio of the thermoelectric voltage to the temperature difference, $S=-\Delta V/\Delta T$.  Based on Fermi liquid theory, Dolgopolov and Gold \cite{dolgopolov11,gold11} obtained the following expression for the diffusion thermopower of strongly interacting 2D electrons in the low-temperature regime:
\begin{equation}
S=-\alpha\frac{2\pi k_B^2m^*T}{3e\hbar^2n_s},\label{eq2}
\end{equation}
where $k_B$ is Boltzmann's constant.  This expression, which resembles the well-known Mott relation for non-interacting electrons, was shown to hold for the strongly-interacting case provided one includes the parameter $\alpha$ that depends on both disorder \cite{fletcher97,faniel07,goswami09} and interaction strength \cite{dolgopolov11,gold11}. The dependence of $\alpha$ on electron density is rather weak, and the main effect of electron-electron interactions is to suppress the thermopower $S$.

\begin{figure}
\centerline{\psfig{file=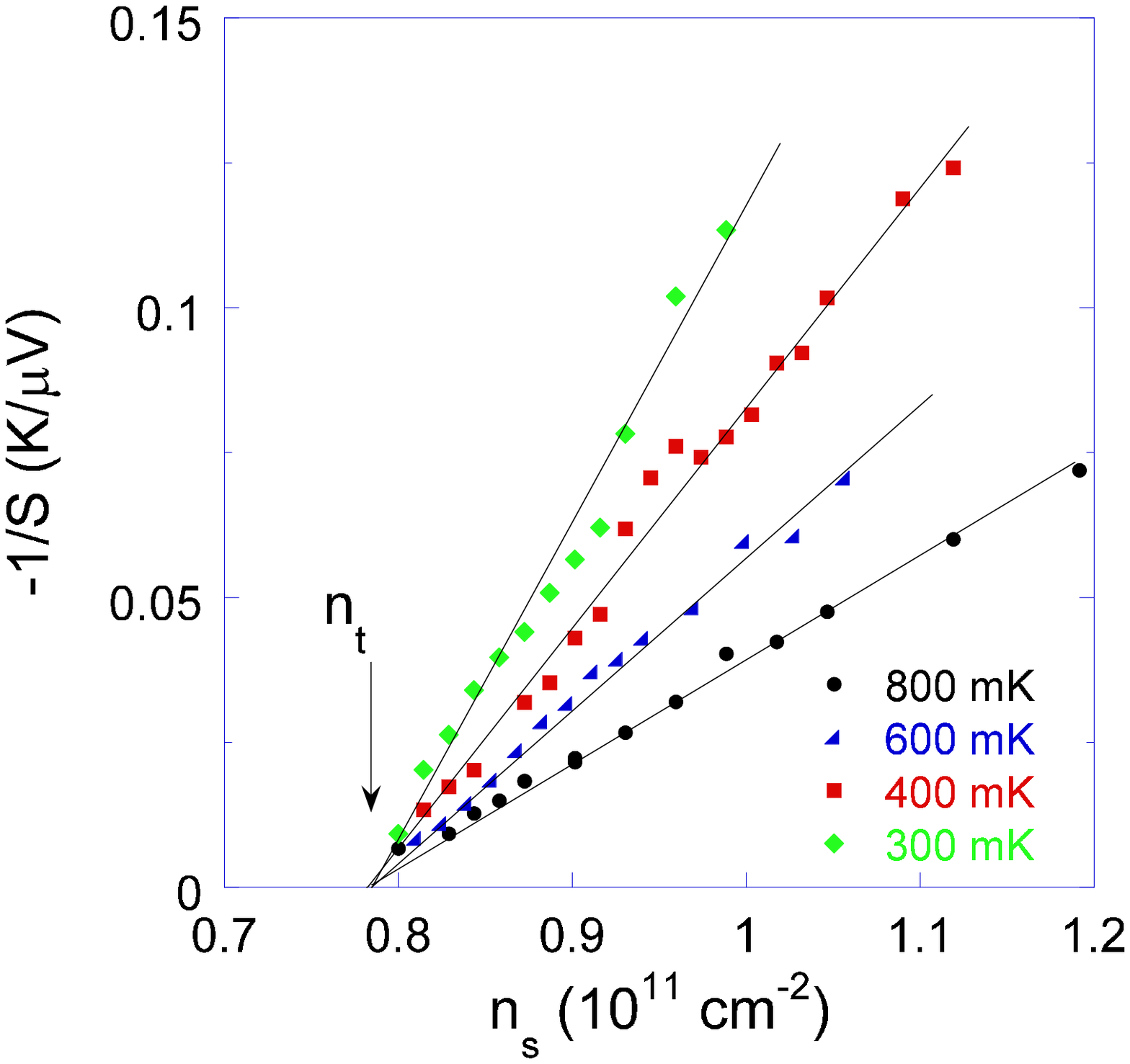,width=3in}\psfig{file=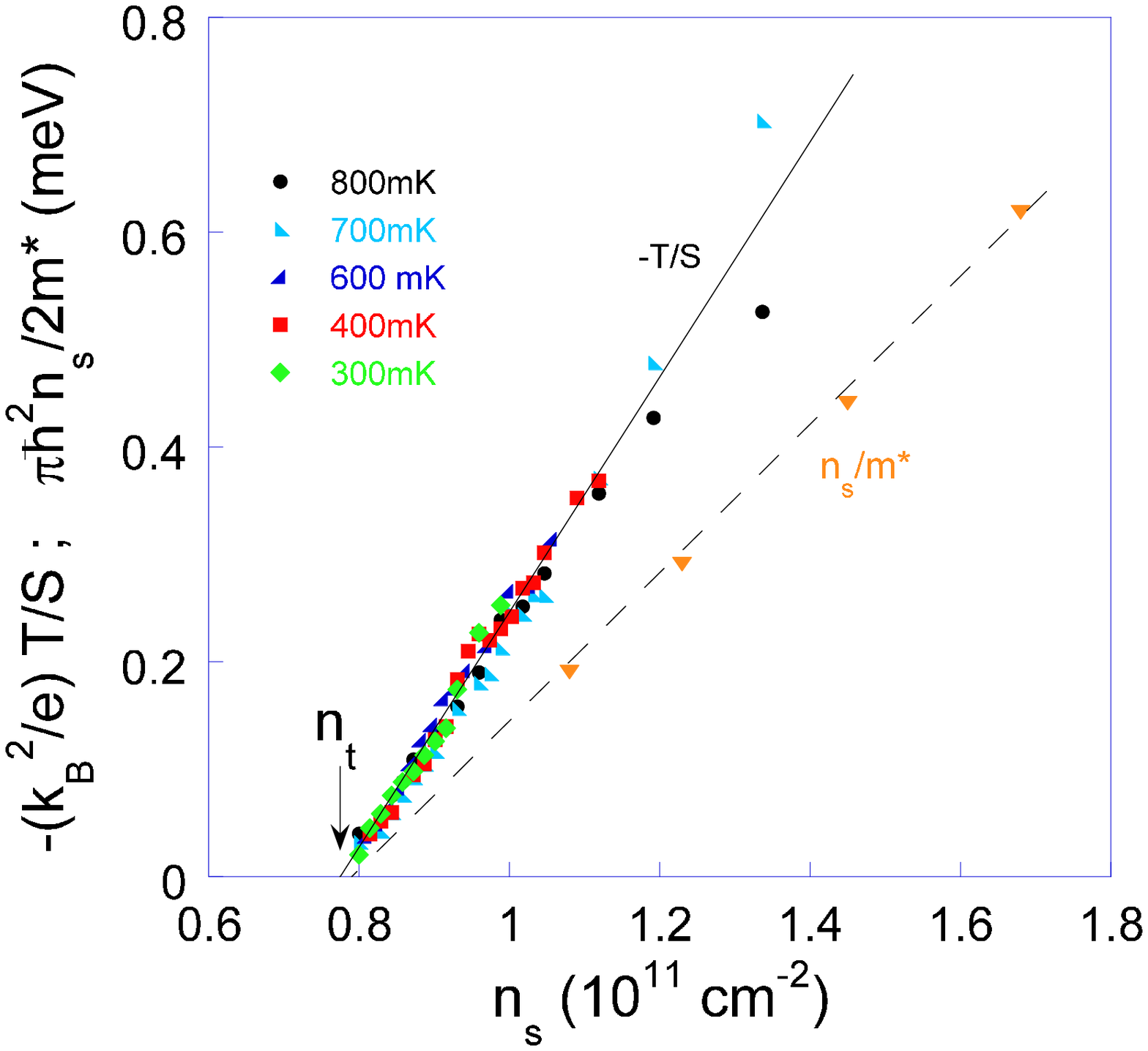,width=3.1in}}
\caption{\label{fig7} Left-hand panel: The inverse thermopower as a function of electron density at different temperatures. The solid lines denote linear fits to the data and extrapolate to zero at a density $n_t$. Right-hand panel: $(-T/S)$ versus electron density for different temperatures. The solid line is a linear fit which extrapolates to zero at $n_t$. Also shown is the effective mass obtained for the same samples by transport measurements \cite{shashkin02}. The dashed line is a linear fit. Adapted from Ref.~\cite{mokashi12}.}
\end{figure}

According to Eq.~(\ref{eq2}), $(-S/T)$ is proportional to $(m^*/n_s)$ and, therefore, the measurements of the thermopower yield the effective mass.  The divergent behavior of the thermopower is evident when plotted as the inverse quantity, $(-1/S)$, versus electron density in Fig.~\ref{fig7}(left-hand panel).  Figure~\ref{fig7}(right-hand panel) shows $(-T/S)$ plotted as a function of $n_s$. The data collapse onto a single curve demonstrating that the thermopower $S$ is a linear function of temperature. In turn, the ratio $(-T/S)$ is a function of electron density $n_s$ of form:
\begin{equation}
(-T/S) \propto (n_s-n_t)^x.\label{eq1}
\end{equation}
Fits to this expression indicate that the thermopower diverges with decreasing electron density with a critical exponent $x=1.0\pm 0.1$ at the density $n_t = 7.8 \pm 0.1 \times 10^{10}$~cm$^{-2}$ that is close to (or the same as) the density for the metal-insulator transition, $n_c \approx 8 \times 10^{10}$~cm$^{-2}$, obtained from resistivity measurements in this low-disorder electron system. The measured $(-T/S)$, shown in Fig.~\ref{fig7}(right-hand panel), decreases linearly with decreasing electron density, extrapolating to zero at $n_t$ and indicating a strong increase of the effective mass by more than an order of magnitude.  The results thus imply a divergence of the electron mass at the density $n_t$: $m^*\propto n_s/(n_s-n_t)$ --- behavior that is typical in the vicinity of an interaction-induced phase transition.

It is interesting to compare these results with the effective mass obtained earlier for the same samples. As seen in Fig.~\ref{fig7}(right-hand panel), the two data sets display similar behavior. However, the thermopower data do not yield the absolute value of $m^*$ because of uncertainty in the coefficient $\alpha$ in Eq.~(\ref{eq2}). The value of $m^*$ can be extracted from the thermopower data by requiring that the two data sets in Fig.~\ref{fig7}(right-hand panel) correspond to the same value of mass in the range of electron densities where they overlap. Determined from the ratio of the slopes, this yields a coefficient $\alpha\approx 0.18$. The corresponding mass enhancement in the critical region reaches $m^*/m_b\approx 25$ at $n_s\approx 8.2\times 10^{10}$~cm$^{-2}$, where the band mass $m_b=0.19m_e$ and $m_e$ is the free electron mass. The mass $m^*\approx 5m_e$ exceeds by far the values of the effective mass obtained from previous experiments. It is important to note that the thermopower experiment includes data for electron densities that are much closer to the critical point than the earlier measurements, and reports much larger enhancement of the effective mass for reasons explained below.

The thermopower as well as the conductivity give a measure of the mass at the Fermi level, while the Zeeman field $B_c$ required to fully polarize the spins measures the mass related to the bandwidth, which is the Fermi energy counted from the band bottom. For $n_s\geq 10^{11}$~cm$^{-2}$, the mass determined by different methods was found to be essentially the same \cite{kravchenko04,shashkin05}. On the other hand, the behavior is different at the densities reached in the experiment in the very close vicinity of the critical point $n_t$ ($n_s <10^{11}$~cm$^{-2}$) where the bandwidth-related mass was found to increase by only a factor $\approx4$. This follows from the fact that the Shubnikov-de Haas oscillations in the dilute 2D electron system in silicon reveal one switch from cyclotron to spin minima (the ratio of the spin and cyclotron splittings reaches $\approx1$) as the electron density is decreased \cite{kravchenko00a}, the spin minima surviving down to $n_s\approx n_c$ and even below \cite{iorio90}. In effect, while the bandwidth does not decrease appreciably in the close vicinity of the critical point $n_t$ and the effective mass obtained from such measurements does not exhibit a true divergence, the thermopower measurements yield the effective mass at the Fermi energy, which does indeed diverge.

A divergence of the effective mass has been predicted by a number of theories: using Gutzwiller's theory \cite{dolgopolov02}; using an analogy with He$^3$ near the onset of Wigner crystallization \cite{spivak03,spivak04}; extending the Fermi liquid concept to the strongly-interacting limit \cite{khodel08}; solving an extended Hubbard model using dynamical mean-field theory \cite{pankov08}; from a renormalization group analysis for multi-valley 2D systems \cite{punnoose05}; by Monte-Carlo simulations \cite{marchi09,fleury10}. Some theories predict that the disorder is important for the mass enhancement \cite{punnoose05,marchi09,fleury10}. In contrast with most theories that assume a parabolic spectrum, the authors of Ref.~\cite{khodel08} stress that there is a clear distinction between the mass at the Fermi level and the bandwidth-related mass. In this respect, our conclusions are consistent with the model of Ref.~\cite{khodel08} in which a flattening at the Fermi energy in the spectrum leads to a diverging effective mass. This Fermi liquid-based model implies the existence of an intermediate phase that precedes Wigner crystallization.

There has been a great deal of debate concerning the origin of the interesting, enigmatic behavior in these strongly interacting 2D electron systems. In particular, many have questioned whether the change of the resistivity from metallic to insulating temperature dependence signals a phase transition, or whether it is a crossover. We close by noting that, unlike the resistivity which displays complex behavior that may not distinguish between these two scenarios, we have shown by measurements of the spin susceptibility, effective mass, and thermopower that the 2D electron system in low-disordered silicon MOSFETs behaves critically at a well-defined density, providing clear evidence that this is a transition to a new phase at low densities. The next challenge is to determine the nature of this phase.

\end{document}